# Simulating the transport of dissolved ammonium-nitrogen and phosphate-phosphorous to streams using artificial neural network (ANN)

Amir Sedaghatdoost[1]

[1]PhD student, Department of Biological and Agricultural Engineering, Texas A&M University, College Station, TX, 77843, USA.

**Abstract**

The export of agricultural fertilizers to streams results in untold harms to downstream water quality and thus ecosystem. Understanding and simulating the export of these chemicals is the first step to control their transport to streams. In this paper, the export of ammonium and phosphate, two major contaminants resulted from agricultural activities, were simulated using artificial neural network (ANN) and daily sediment and runoff data. Results indicated that ANN could simulate the trend of the export of these chemicals to streams successfully; however, it could not simulate the extreme events well. This might be due to the complexity of the transport of ammonium and phosphate, which contains many different processes that could not be captured in daily sediment and runoff data.

## 1. Introduction

Human activities such as intensive agriculture, often alter surface water chemistry. Ammonium-nitrogen, phosphate-phosphorous, and sediment losses are only a few possible consequences of soil erosion and biochemical applications. Such contaminants are the prime food supply of filter feeders, regulate the water turbidity, limit light penetration and photosynthesis, accelerate the degradation of lakes and rivers, affect biodiversity, and promote the occurrence of water-related diseases. Because of these facts, non-point source pollution at the basin scale is an international research focus, in which nitrate contamination caused by the excess application of nitrogen fertilizer in farmlands is one of the key environmental issues. The excess nitrogen fertilizer might transport into the surface water via drainage systems (Sedaghatdoost et al. 2019; Sedaghatdoost et al. 2018) or cause groundwater contamination by penetrating into soil layers below the crop root zone and ultimately leaching into the aquifer (Mokari et al., 2019). Simple tools are needed to support decision-making and to develop best-management practices to tackle the stream pollution issues since the processes controlling water chemistry in a catchment are very complex.

A simple tool that can be used in studies of nitrate movement to streams is the artificial neural network (ANN). ANNs have evolved from intelligent computer-based systems mimicking the human brain to non-linear mathematical algorithms that map a set of input variables to a set of output variables. As such, ANNs need no explicit information on the processes causing the response. They are entirely data-driven, which often allow reducing the magnitude of the necessary information to build them, in opposition to standard physically-based models. The disadvantage of ANNs is that they provide no physical or chemical information about the occurring processes—it is challenging to decipher neural weights and biases. Nonetheless, there has been a growing trend for the experimentation of ANNs in the hydrologic and water quality domains (Araghinejad et al., 2018; Chau, 2006). Reports of ANN applications simulating nitrate-nitrogen losses are given by Sharma et al. (2003), Suen and Eheart (2003), and May and Sivakumar (2008), while reports of ANN applications simulating suspended sediment losses are given by Nour et al. (2006), and Zhu et al. (2007). The purpose of the paper is to assess the application of ANN in the simulation of daily ammonium-nitrogen and phosphate-phosphorous from a small agricultural sub watershed to a stream.

## 2. Materials and Methods

### *Study Area*

All of the studied watersheds were located at the USDA-ARS Grassland, Soil and Water Research Laboratory near Riesel, Texas (Fig. 1). The study area has Vertisol soils that shrink and swell rapidly due to the changes in soil moisture content. Although the hydraulic conductivity of these soils is very low, preferential flow due to cracks plays a significant low in water infiltration (Harmel et al., 2009).

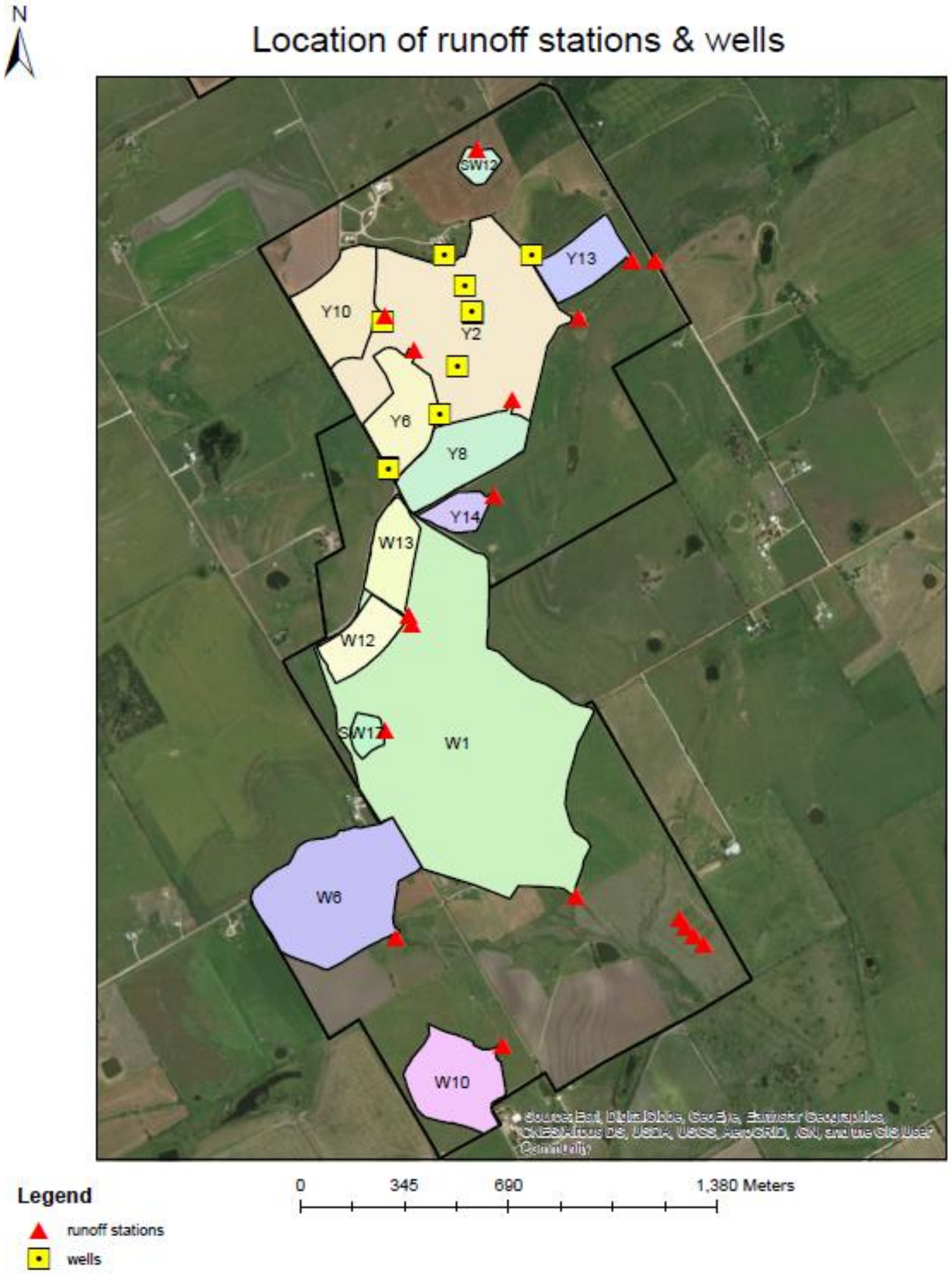

**Fig. 1- The location of the study area**

***Artificial Neural Network***

Artificial neural network modeling of a process demands two operations: training and thereafter testing (Fig. 2). Training involves the optimization of the connected weights through the minimization of a specific function. The training step is a significant period of a neural network, in which an input is introduced to the network together with the desired target. The weights and biases are adjusted iteratively so that the network attempts to produce the desired output. The data were divided into a training set of 46 data points and a testing set of 55 points (Fig. 3). The training was halted as soon as the mean square error was unchanged and was significantly small (in the order of 0.001) for the number of epochs (iterations) to prevent overfitting.

***Statistical Evaluation***

To evaluate the simulation results of the ANN technique, $R^2$ and root mean square error was used.

$$RMSE = \sqrt{\frac{\sum(P_i - O_{i)}}{n}}$$

Where $P_i$ is the predicted values, $O_i$ is observed values, and $n$ is the number of observations.

## 3. Results and Discussions

Results indicated that ANN could simulate the trend of the export of these chemicals to streams successfully; however, it could not simulate the extreme events well (Fig. 4). It is because the transport of ammonium and phosphate contains many complex processes that could not be captured in daily sediment and runoff data. Overall, the evaluation criteria showed that ANN could simulate phosphate better than ammonium; however, it is clear that if the extreme events were excluded, the results of ammonium would be much better than phosphate (Fig. 5).

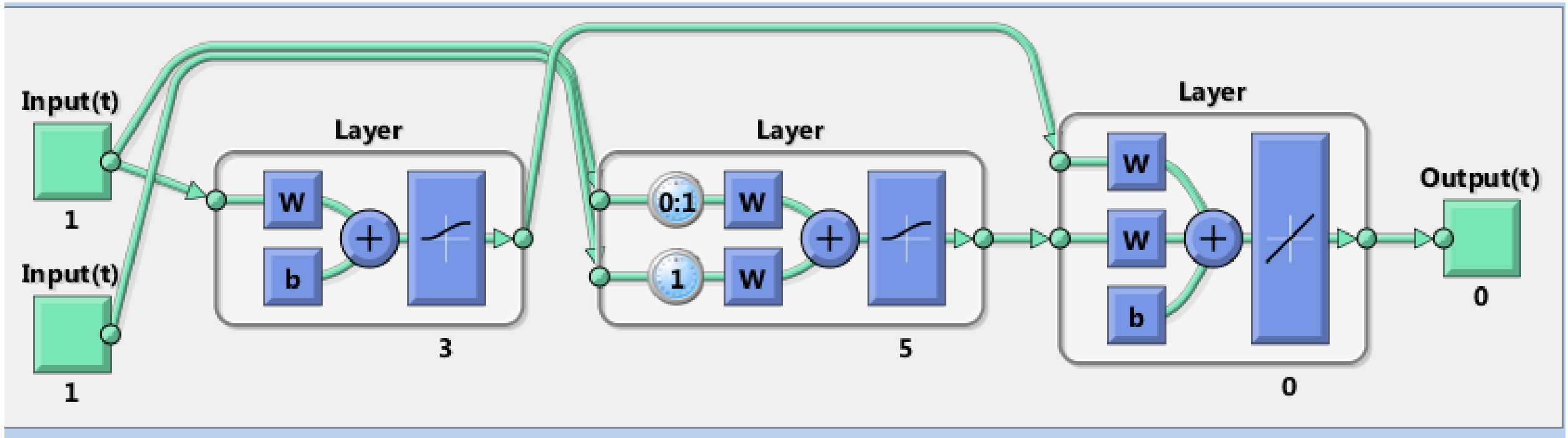

**Fig. 2- The diagram of applied artificial neural network**

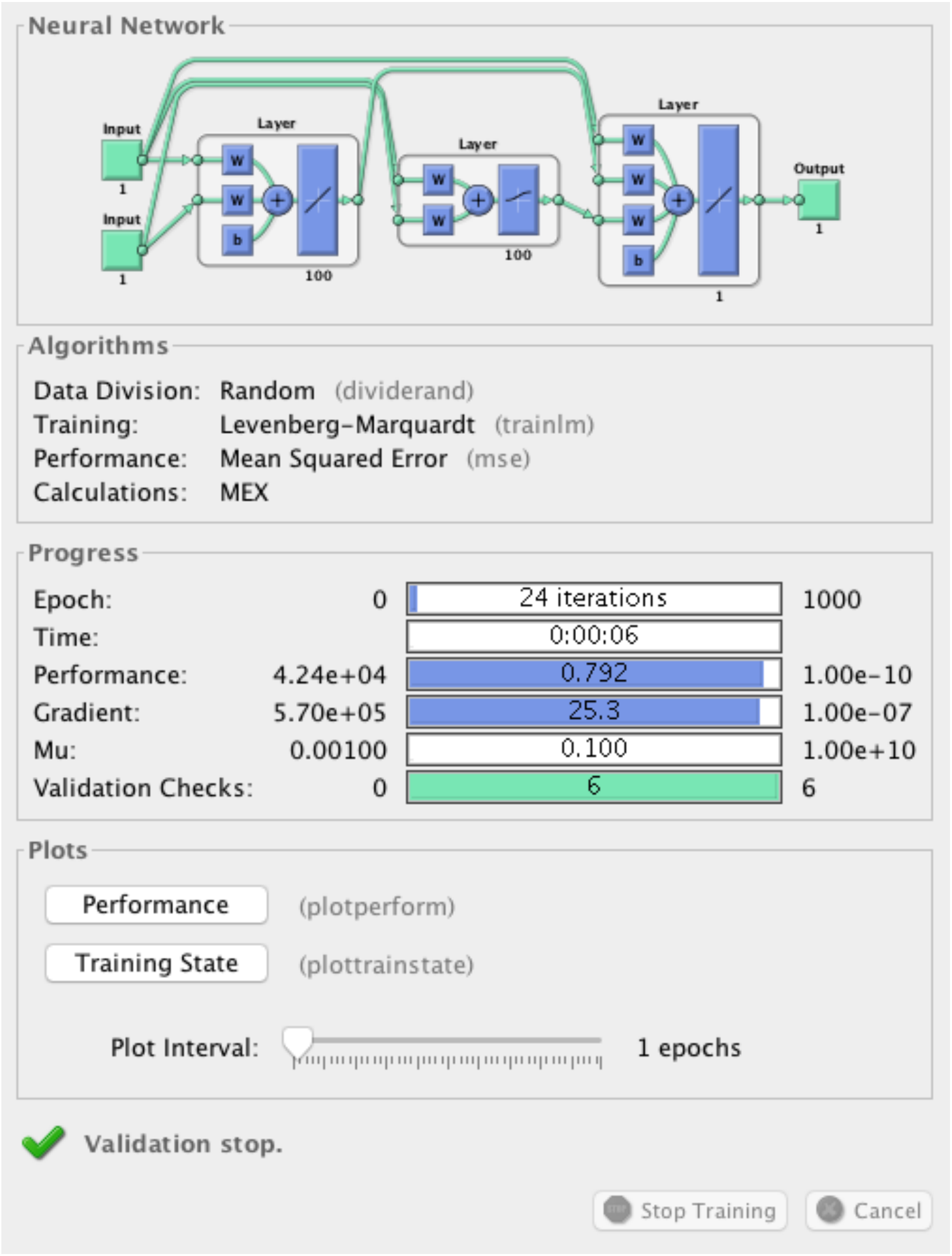

**Fig. 3- A summary of properties associated with applied ANN**

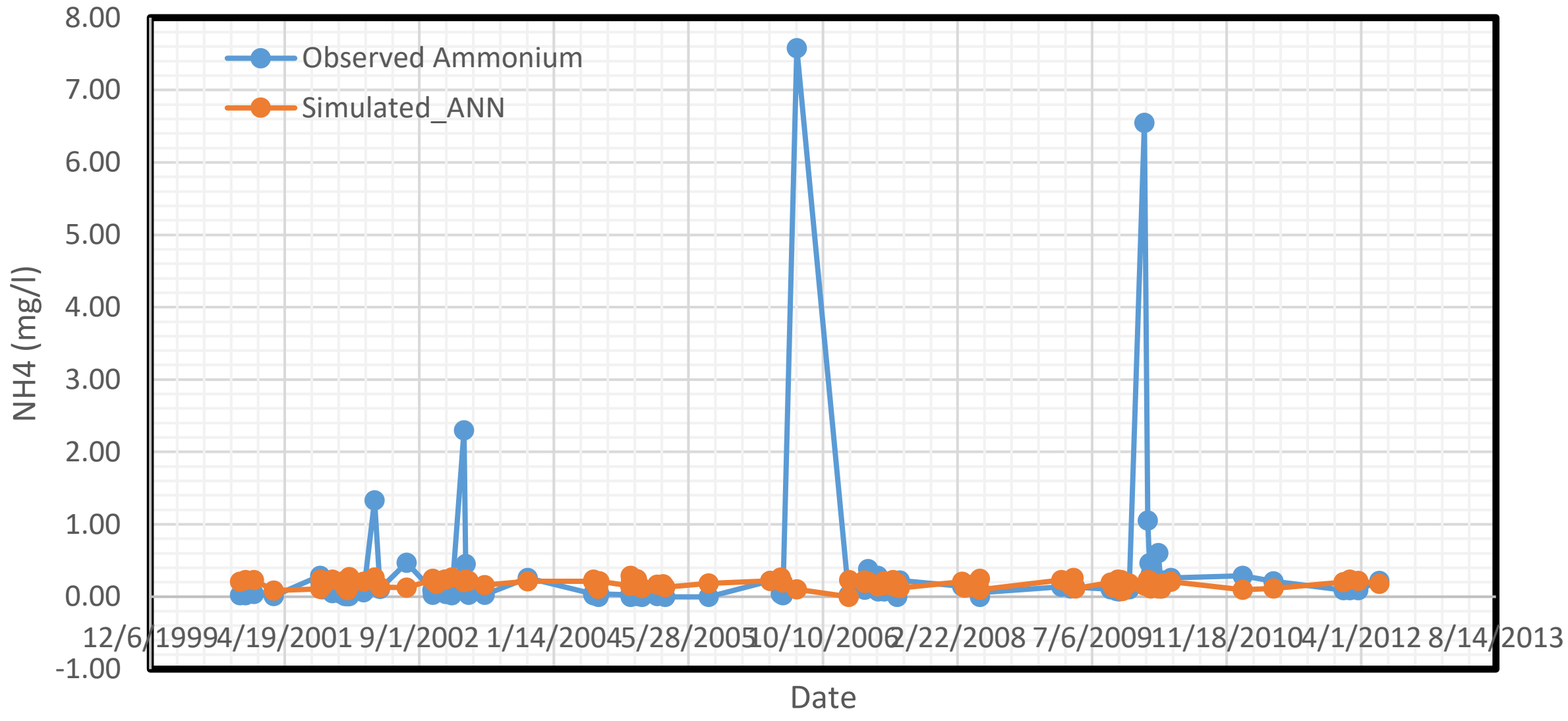

**Fig. 4- A plot of observed and simulated values of ammonium-nitrogen versus time**

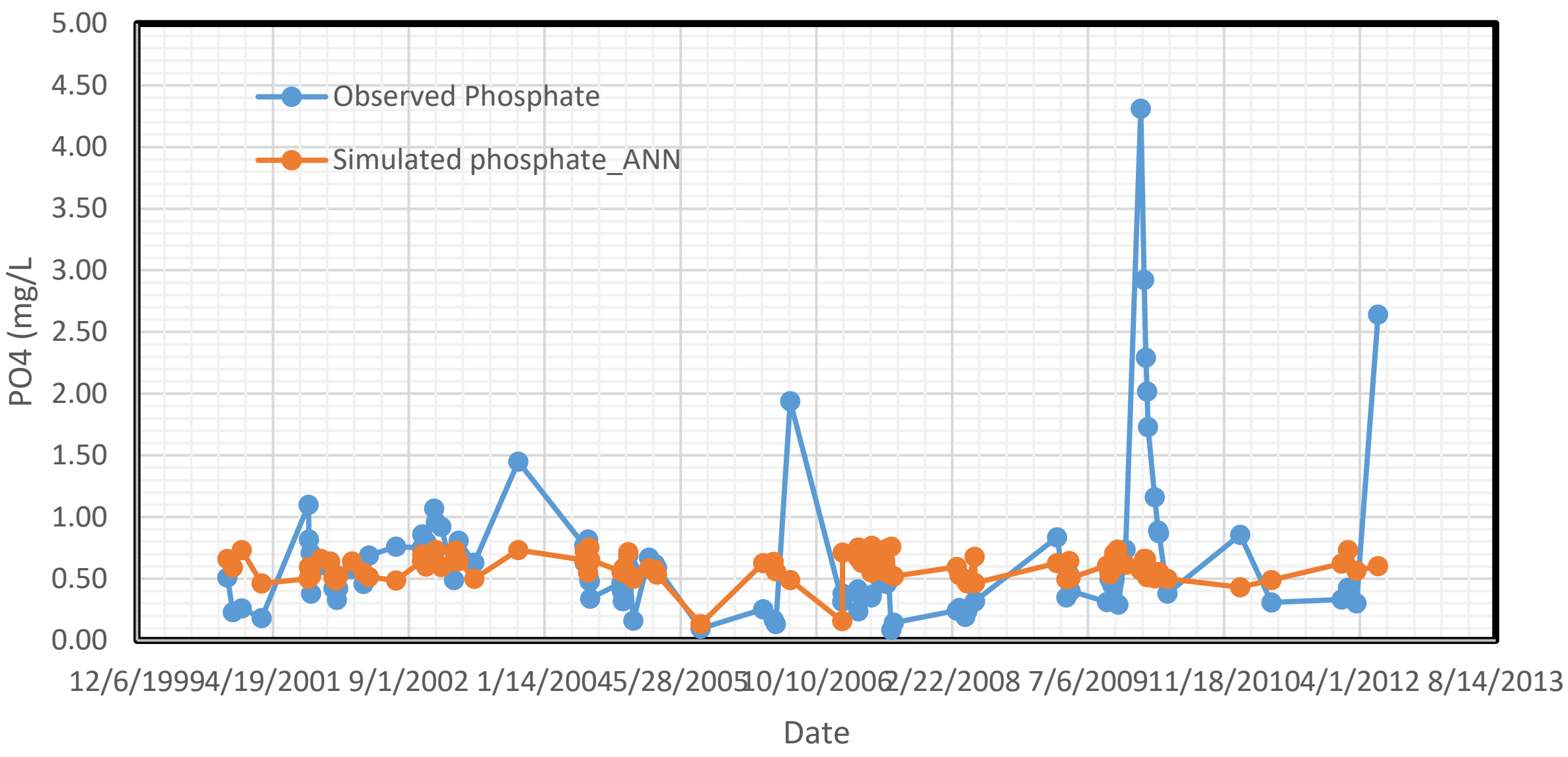

**Fig. 5- A plot of observed and simulated values of phosphate-phosphorous versus time**

Table 1 also indicates that the error in predicting ammonium and phosphate is relatively low; however, $R^2$ value is not high. This shows that ANN can apply to simulate the transport of ammonium and phosphate into stream; but, cannot be used to find the trend of the changes in these two chemicals.

**Table 1- Evaluation criteria for the simulated values of ammonium and phosphate**

| **Contaminant** | **$NH_4$** | **$PO_4$** |
|---|---|---|
| **$R^2$** | 0.02 | 0.01 |
| **RMSE (mg/L)** | 1.04 | 0.63 |

## 4. Conclusion

The results indicated that ANN could be applied as a useful for predicting the trend of contaminant export to streams using runoff and sediment data; however, for obtaining more accurate results, adding more information such as fertilizer application or soil nitrogen and phosphorous content should be added to the ANN technique.